# Why Are Verdazyl Radicals Non-Emissive? An Experimental and Computational Study

*Alexandre Malinge, Pierre-Luc Thériault and Stéphane Kéna-Cohen\**

Department of Engineering Physics, École Polytechnique de Montréal, PO Box 6079, succ. Centre-Ville, Montreal QC H3C 3A7, Canada

ABSTRACT

Verdazyl radicals are a versatile class of air-stable organic radicals used in various applications, especially for their magnetic properties. Despite the development of a wide range of verdazyl derivatives, however, they are all non-emissive. To investigate the reasons behind this and to understand the excited-state dynamics of verdazyls, we combine steady-state and femtosecond pump-probe spectroscopy with quantum chemical calculations. In the carbazole-substituted 2,4,6-triphenylverdazyl (TPV-Cz) , we observe ultrafast internal conversion of the first excited state on a timescale of $0.5 \pm 0.1$ ps, followed by vibrational relaxation with a lifetime of $3.7 \pm 0.4$ ps. Spin-flip time-dependent density functional theory calculations reveal that the sub-picosecond non-radiative decay comes from a low-energy conical intersection between the $D_1$ and $D_0$ states, driven by an out-of-plane distortion of the verdazyl ring. This distortion is observed and remains energetically accessible in the isolated verdazyl ring, in 2,4,6-triphenylverdazyl, and in TPV-Cz. This shows that the conical intersection geometry is a recurring feature across different types of verdazyl derivatives and explains why all verdazyls are non-emissive despite different functionalization. Our results provide a mechanistic understanding of the photophysical properties of verdazyl radicals and offers a pathway for the future design of emissive verdazyl derivatives.

1. **Introduction**

Luminescent organic radicals have recently attracted growing interest as an alternative class of emitters for optoelectronics and quantum technologies. In the vast majority of closed-shell organic molecules, the lowest excited state is a triplet. Given that radiative transitions from the triplet to the singlet ground state are spin-forbidden, organic light-emitting devices must use triplet harvesting schemes or phosphorescent emitters to harness electrically generated triplet excitons. Even in these cases, however, the long lifetime typical of triplet excitons reduces the efficiency at high brightness due to intermolecular processes involving triplet excitons[1–3]. In contrast, organic radicals have doublet ground and excited states with a spin-allowed optical transition[4,5]. For this reason, the theoretical internal quantum efficiency of organic light-emitting diodes containing luminescent organic radicals can be 100%, without the drawbacks posed by architectures that rely on harnessing triplets. Finally, the unpaired electron of radical can also be used as a spin qubit for storing and manipulating quantum information[6–12] and luminescent radicals allow for the possibility of efficient optical spin read out.

Luminescent radicals based on triarylmethyl, especially tris(2,4,6-trichlorophenyl)methyl (TTM) derivatives, have been the most studied. In 1944, Lewis et al. reported the first evidence of fluorescence at 90 K in triarylmethyl radicals[13] and room-temperature fluorescence in perchlorotriphenylmethyl radical was observed with a quantum yield of 1.5 % in 1987[14]. These radicals, however, are not photostable and undergo a fast cyclization upon irradiation.

A major milestone was reached in 2006 with the design of a TTM radical coupled to a carbazole donor (TTM-Cz)[15]. In TTM, molecular symmetry leads to degenerate HOMO–SOMO and SOMO–LUMO excitations, which cancels the transition dipole moment. The introduction of the carbazole substituent breaks this symmetry and induces a charge-transfer state[16]. This results in a significant increase of the photoluminescence quantum yield from 3 % for TTM to 53 % for TTM-Cz in cyclohexane. It also changes the excited-state decay pathways and limits photocyclization of the TTM core. The half-life of the emission intensity under continuous irradiation increases from 63 s for TTM to $3.3 \times 10^3$ s for TTM-Cz[17]. TTM-Cz and related radicals have shown excellent performance in OLEDs[18,19], even at high current densities and inspired further derivatives with improved photostability and quantum efficiency[20–25]. To date, the development of luminescent radicals has been mostly restricted to TTM-based structures. This places strong limitations on the accessible range of photophysical and electronic properties for such compounds, limiting potential use cases.

New classes of luminescent radicals are challenging to develop because most radical species are unstable and react quickly with oxygen and moisture[26]. Verdazyls, however, are one of the most important exceptions to this rule. Verdazyls are highly stable and can be stored for months or even years without any sign of degradation. This stability is due to the strong delocalization of the unpaired electron over four nitrogen atoms in the cycle[27]. Owing to their excellent stability, verdazyls have been widely studied for various applications. They are extensively used as molecular magnets[27] due to their spin properties and tunable ferromagnetic or antiferromagnetic exchange interactions[28] and have also been explored as building blocks in organic electronic and spintronic devices[29,30]. Additionally, their ability to form charge-transfer complexes opens possibilities in organic redox flow batteries and catalysis because they exhibit chemically stable open-shell states and reversible multi-electron redox processes over a wide potential window [31–35]. These applications underline the versatility of verdazyl radicals. Recently, it has been shown that restraining the molecular motion of the verdazyl could be helpful to promote anti-Kasha emission ($D_2 \rightarrow D_0$) from moieties covalently coupled to the verdazyl ring[36,37]. However, their photophysical properties remain largely unexplored and despite the development of many verdazyl derivatives, none show strong luminescence.

In this work, we unveil why verdazyl radicals are dark through a combination of experiment and quantum chemical calculations. We perform density functional theory (DFT), linear response time-dependent DFT (LR-TDDFT), and spin flip TDDFT (SF-TDDFT) calculations to analyze the electronic structure and understand the excited-state dynamics. We study three molecules to evaluate how substitutions influence photophysical properties: the 2,4,6-triphenylverdazyl (TPV), the isolated verdazyl ring (VR), and a newly synthesized verdazyl coupled to a carbazole donor group (TPV-Cz). The photophysical properties of TPV-Cz are characterized using absorption, transient absorption and photoluminescence spectroscopy.

## 2. Results and discussion

### 2.1. Energetic Impact of Phenyl Group Motions in 2,4,6-Triphenylverdazyl

One of the simplest and most studied verdazyl is TPV[38] (Table 1). Its structure and synthesis are straightforward and the three phenyl substituents provide steric protection to the radical[39,40]. As a result, a wide range of verdazyl derivatives have been developed based on this structure.

In 2015, Weinert *et al*. investigated the excited-state dynamics of TPV with femtosecond transient absorption[41]. They observed ultrafast $D_1$ to $D_0$ internal conversion within a few hundred femtoseconds, followed by vibrational relaxation over a few picoseconds. In 2024, ultrafast non-radiative decay was also reported in verdazyl diradicals[42]. These studies suggested that the ultrafast decay must arise from a conical intersection (CI) between $D_1$ and $D_0$[43]. This process was attributed to motion of the phenyl rings around the verdazyl core. However, there has been no computational confirmation that a CI truly exists in verdazyl, let alone to find its geometry.

To explore this problem, we used DFT and TDDFT to study electronic transitions, orbital configurations and geometries. SF-TDDFT calculations were performed to explore the potential energy surfaces (PES) of the $D_0$ and $D_1$ states. Standard LR-TDDFT cannot be used around CIs as it treats excited states as perturbations of the ground state. This difference in treatment between ground and excited states leads to a dimensionality problem near degeneracies and prevents it from capturing the correct topology of the CI[44,45]. SF-TDDFT overcomes this limitation by using a high-spin state (in our case, a quartet) as a reference and describes both the ground and excited states as spin-flip excitations[46,47]. Multiconfigurational methods like CASSCF can also locate CIs, but they are limited to small systems due to the choice of a limited number of orbitals (active space) and the computational cost[43,48,49]. In contrast, SF-TDDFT does not require an active space and remains computationally feasible for larger open-shell molecules like TPV. This is therefore the most practical choice for locating CI in our molecules. The choice of functional was guided by literature and the basis set was selected following a basis set convergence test of the total energy (Figure S1 and S2).

To evaluate the hypothesis that the CI arises from distortions of the phenyl rings, we investigate whether such motions can indeed bring the $D_0$ and $D_1$ states into degeneracy. We optimize the $D_0$ and $D_1$ geometries of TPV and analyze the structural changes associated with the electronic excitation to $D_1$ (Table 1). The $\widehat{NNNN}$ dihedral remains 0.0° in both states and the main relaxation to $D_1$ involves a partial planarization of the structure, associated with torsion and tilting of the phenyls around the $\widehat{NNNN}$ plane. These structural changes are particularly pronounced for the N-phenyl groups with a variation of 6.2° and 9.1° for tilting and torsion angles, respectively. The C-phenyl substituent is less affected, with variations of 5.5° and 2.2° for tilting and torsion angles.

We also compute the PES of $D_0$ and $D_1$ states along these motions. For both N-phenyl (Figure 1a) and C-phenyl substituents (Figure 1b), the energy variation is relatively small along the torsional coordinate, with a maximum deviation of 0.5 eV at the optimized tilting angle of ~170°. Moving the tilting angle by ± 30° around this equilibrium geometry induces only modest changes of 0.5 eV for N-phenyls and 0.7 eV for C-phenyls. Beyond this range, the energy of both $D_0$ and $D_1$ rises sharply. Overall, no significant convergence between the $D_0$ and $D_1$ PES is observed.

We also compute the symmetric deformation of the N-phenyl groups of TPV (Figure 1c) to reproduce the structural change observed between the $D_0$ and $D_1$ geometries. As expected, the interconversion energy barrier is slightly higher for $D_0$ (0.3 eV) than for $D_1$ (0.1 eV), consistent with the more planar geometry of the $D_1$ minimum. However, the energetic variation remains small, and the two PESs are nearly identical.

This limited influence of the phenyl substituents on the energy gap can be understood by analyzing the electronic structure of TPV. Natural transition orbital (NTO) analysis (Figure 2a) reveals that the first excited state mainly arises from a HOMO → SOMO transition (94 % of the total excitation). The SOMO is mostly delocalized over the four nitrogen atoms with some extension in the N-phenyl substituents (Figure 2b). It is mostly an antibonding orbital across the nitrogen atoms of the verdazyl ring with some non-bonding character on the N-phenyl groups. No delocalization is observed toward the C-phenyl substituent due to the presence of a nodal plane at the connecting carbon atom that prevents conjugation in this direction. The HOMO is also primarily localized in the verdazyl ring but with extended delocalization to phenyl groups. The transition density difference (Figure 2c) confirms that the electronic redistribution upon excitation is almost entirely confined to the verdazyl ring with minimal change on the phenyl substituents. This spatial confinement of the excitation explains why structural distortions of the phenyl groups have only little effect on the $D_0 - D_1$ energy gap.

These results indicate that isolated phenyl motions around the verdazyl ring are unable to lead to a CI. Only extreme distortions induce significant energetic changes in both PES, but even in such cases, the energy gap between $D_0$ and $D_1$ remains far too large to enable a crossing. In contrast, the spatial confinement of the excitation suggests that the energy gap should be highly sensitive to structural distortions of the central verdazyl ring.

Additional 2D PES scans of tilting and torsion angles at the $D_0$ optimized geometry and 3D PES of simultaneous deformations of both C-phenyl and N-phenyl groups are provided in the Supporting Information (Figure S3 – S5).

### 2.2. CI Pathway via Out-of-Plane Distortion of the Verdazyl Ring

In unsaturated rings, out-of-plane distortions are common and have been previously associated with excited-state relaxation pathways[50,51]. For instance, in nitrogen-containing heterocycles such as adenine, pyramidalization of amino groups has been shown to drive the molecule towards CI geometries[52]. We therefore explore analogous distortions in the verdazyl ring (VR) by scanning two internal coordinates: the dihedral angle between the four nitrogen atoms $\widehat{NNNN}$ and the angle involving the adjacent N–H group $\widehat{NNNH}$ to probe out-of-plane distortion of the ring. In this simplified model, the phenyl groups are removed to focus on distortions of the verdazyl ring and to reduce computational cost.

The PES of the $D_0$ state (Figure 3a) of VR shows a minimum at a planar geometry with $\widehat{NNNN}$ of 0.0°. Upon deviation of $\widehat{NNNN}$ by more than ~10°, a steep rise in the $D_0$ surface is observed. In contrast, the $D_1$ surface is relatively unaffected by deviation between 0 and -18°. The $D_1$ minimum is characterized by a deviation from planarity of -11.5°. This distortion follows a symmetric out-of-plane mode, where the two diagonal nitrogen atoms are displaced in the same directions. This is coupled with an antisymmetric displacement of the N–H groups, which move in opposing directions above and below the plane of the ring.

The $D_1$ geometric differences between VR and TPV minima are mainly observed at the N-phenyl substituents. In TPV, steric interactions between the phenyls and the $CH_2$ unit prevent the antisymmetric displacement and lead to a symmetric motion of the N-phenyl groups.

These motions are consistent with the redistribution of electron density (Figure S6), which creates repulsion between adjacent nitrogen atoms and attraction between diagonal ones.

The PES scan and conical intersection optimization reveals a CI located at –16.3° for $\widehat{NNNN}$ and 279.4° for $\widehat{NNNH}$. The norm of the nonadiabatic coupling matrix element (NACME) at this geometry reaches a large value of 11.99 bohr$^{-1}$, which is consistent with the presence of a conical intersection between $D_0$ and $D_1$. The variation of the NACME along the minimum energy path (MEP) is shown in Figure S7. This geometry is characterized by the out-of-plane deviation of one nitrogen atom and its N–H group. This distortion limits the strong delocalization of the unpaired electron across the four nitrogen atoms, leading to a significant destabilization of the $D_0$ state. In contrast, upon excitation in the verdazyl ring, the unpaired electron in $D_1$ occupies an orbital that is strongly localized on the sp² carbon and on the two non-adjacent nitrogen atoms (Figure 4). As a result, the out-of-plane distortion has a much weaker influence on the stabilization of the unpaired electron in $D_1$.

The relaxation from the vertical excitation at the $D_0$ minimum to the CI follows a barrierless pathway (Figure 3b). Moreover, the MEP between $D_1$ and $D_0$ reveals another CI between $Q_1$ and $D_2$ that follows the same geometric feature. This easily accessible crossing provides an efficient non-radiative decay channel for $D_1$, which explains the ultrafast excited-state decay observed experimentally and the non-emissive character of verdazyls.

### 2.3. Effect of Carbazole Substitution on the Excited-State Dynamic

The introduction of an electron-donating group in radicals generally changes the nature of the transition by inducing a strong localization of the HOMO on the donor. This strategy has been successfully used in TTM to create a charge-transfer state and promote fluorescence[16]. Following this strategy, we couple a carbazole moiety to TPV scaffold to create a push–pull system and investigate its impact on the excited-state dynamics of verdazyl.

The carbazole-substituted verdazyl (TPV-Cz) was synthesized following the classical procedure (Scheme 1). The synthesis involves three main steps: (i) condensation of phenylhydrazine with the appropriate aldehyde to yield the hydrazone, (ii) diazo coupling with an aryl diazonium salt to form the corresponding formazan, and (iii) cyclization with formaldehyde under acidic conditions followed by spontaneous air oxidation to afford the verdazyl radical. Full experimental details, including the synthetic procedure, NMR, EPR and mass spectrometry data, are available in the Supporting Information.

To characterize the compound and its photophysical properties, we perform a series of spectroscopic analyses, including UV–VIS absorption, photoluminescence and transient absorption.

The absorption spectrum of TPV-Cz in toluene shows a relatively small broad band centered at 730 nm (Figure 5a, left), which is a characteristic signature of a HOMO-SOMO transition in verdazyls, with two stronger bands at 425 nm and 330 nm. Upon photoexcitation at 710 nm, only a very weak emission is observed at 820 nm in 5 wt.% and 1 wt.% doped PMMA thin films at 4 K (Figure 5a, right and Figure S8). Attempts to determine the PLQY at 298 K were unsuccessful due to the extremely weak photoluminescence of the compound and is estimated to be less than 0.01%.

To understand the excited-state dynamics of TPV-Cz, we performed femtosecond pump – probe spectroscopy on toluene solutions containing TPV-Cz and PMMA thin films doped at 5 and 10 wt.%. . Absorption spectra were compared before and after irradiation to confirm the absence of photodegradation and no pump power and concentration dependance were observed (Figure S9).

Upon excitation at 750 nm, we observe two excited-state absorption (ESA) bands centered at 480 nm and 900 nm, both in solution (Figure 5b) and in thin films (Figure S10). These bands appear within one picosecond (Figure S11) and show a rise time longer than the pump pulse duration and the instrument response. This excludes an instrumental effect and suggests that the ESA bands arise not directly from the $D_1$ state but from a state populated through ultrafast internal conversion within the first picosecond.

After this rise, the signal decays monoexponentially and fitting yields an average lifetime of 4.0 $\pm$ 0.1 ps in solution (Figure 5c). In PMMA (Figure S12), a slower decay of 18.0 $\pm$ 0.5 ps is attributed to restricted conformational relaxation in the solid matrix. The absence of a GSB band in our system is consistent with observations in other verdazyl-based systems. In TPV, the ground-state bleach was reported to be six times weaker than the ESA signal[41]. This observation was attributed to the presence of an additional absorption band that compensates the ground-state depletion and masks the expected bleach. In our case, the presence of an additional ESA band in the near-infrared region further increases the spectral overlap with the GSB band and prevents any clear observation of the bleach.

Singular value decomposition of the transient absorption data indicated the presence of two significant components (Figure S13). A global fit using a three-state kinetic model provided the best agreement with the experimental data: an initial sub-picosecond population transfer ($0.5 \pm 0.1$ ps) between two excited states, followed by a monoexponential decay ($3.7 \pm 0.4$ ps) to the ground state. The ultrafast kinetics, lack of concentration effects, and similarity of behavior in both solution and solid state make electron or energy transfer mechanisms unlikely. This sub-picosecond timescale is characteristic of a relaxation pathway involving a CI between $D_1$ and $D_0$, while the second is attributed to vibrational relaxation in the $D_0$ state.

This interpretation is supported by the analysis of the effect of substitution on the CI geometry. We compute the energies of the $D_0 \rightarrow D_1$ vertical transition and the energy of the CI (Table 2) for VR, TPV and TPV-Cz. We also report the geometry of $D_0$ and CI through $\widehat{NNNN}$ and tilting $\widehat{NNNR}$ dihedral angles.

We found a CI in both TPV and TPV-Cz with structural and energetic features comparable to VR. The CI is still characterized by a loss of planarity in the ring and is more pronounced compared to VR, with the $\widehat{NNNN}$ dihedral angle reaching -33.8° in TPV and -33.0° in TPV-Cz. The $\widehat{NNNR}$ dihedral angle shows a large variation of over +100° in all molecules at the CI with a smaller distortion observed in substituted verdazyls.

The energy of the CI is similar upon substitution (1.68 eV in VR, 1.65 eV in TPV and 1.69 eV in TPV-Cz) but remains well below the vertical excitation energy and is still readily accessible. This explains why the ultrafast non-radiative decay is observed in both derivatives.

Moreover, the orbital analysis of TPV-Cz reveals that the donor group substitution only leads to a partial delocalization of the HOMO onto the carbazole (Figure S14), and the resulting charge-transfer character remains too weak to significantly alter the electronic transition density. As the result, the introduction of a carbazole group in TPV does not significantly prevent access to the CI responsible of the ultrafast relaxation decay.

A stronger electron-donating group would be required to further localize the HOMO on the donor and significantly impact the CI. However, since the SOMO remains confined to the nitrogen atoms of the verdazyl ring, this would progressively reduce the spatial overlap between the orbitals, leading to a substantial decrease in oscillator strength. The choice of the donor must therefore carefully balance the ability to perturb the CI with the need to preserve sufficient orbital overlap for emission.

These results show that the CI remains energetically accessible in all derivatives, and that the addition of substituents — even electron-donating groups like carbazole — does not significantly impact its geometry.

Based on these results, two strategies can be considered to design bright verdazyl-based emitters: tuning the nature of the transition towards a CT state to modify the electronic transition density while maintaining sufficient orbital overlap; and structurally constraining the verdazyl core to make out-of-plane motion of the nitrogen highly unfavorable in the $D_1$ state.

These strategies highlight the importance of considering both electronic and geometric factors in the future molecular design of luminescent verdazyl derivatives.

## 3. Conclusion

In summary, our combined experimental and computational results reveal that the dark nature of verdazyl radicals originates from an efficient non-radiative decay pathway involving a conical intersection between the $D_1$ and $D_0$ states. This intersection is driven by an out-of-plane distortion of the verdazyl ring, which destabilizes the $D_0$ state without significantly affecting the $D_1$ state.

To evaluate whether charge-transfer character could influence this process, a carbazole donor group was introduced into the verdazyl scaffold. However, this modification fails to suppress the non-radiative decay. Internal conversion with a sub-picosecond lifetime of $0.5 \pm 0.1$ ps is observed for the $D_1$ state, followed by vibrational relaxation in $D_0$ with a lifetime of $3.7 \pm 0.4$ ps in solution. The carbazole unit does not sufficiently delocalize the transition density away from the verdazyl core and a stronger electron-donating group is required. As a result, it has little effect on the geometry and energy of the conical intersection, and the non-radiative decay pathway remains active.

Overall, these results improve our understanding of the fundamental photophysical mechanisms governing verdazyl radicals and provide design guidelines for the development of new luminescent derivatives. Suppressing the non-radiative decay will require making the deviation from planarity of the verdazyl ring energetically unfavorable, either by constraining the geometry or by modifying the electronic transition density.

**FIGURES**

|  | **N-Phenyl** | | **C-Phenyl** | |
|---|---|---|---|---|
|  | Tilting $\widehat{C_2 N_3 N_4} C_3$ | Torsion $N_3 \widehat{N_4 C_3} C_4$ | Tilting $N_4 \widehat{N_3 C_2} C_5$ | Torsion $N_3 \widehat{C_2 C_5} C_6$ |
| **D$_0$ geometry** | 167.0 | -17.9 | 173.8 | -2.8 |
| **D$_1$ geometry** | 173.2 | -8.8 | 179.3 | -0.6 |

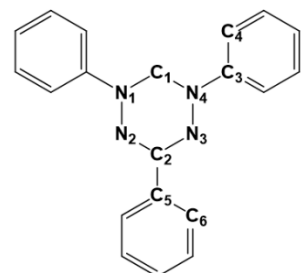

**Table 1.** Relevant change in dihedral angles (°) between the optimized D$_0$ and D$_1$ geometries of TPV. The molecular structure with atom labels used throughout this work is shown to the right. Hydrogen atoms are omitted for clarity.

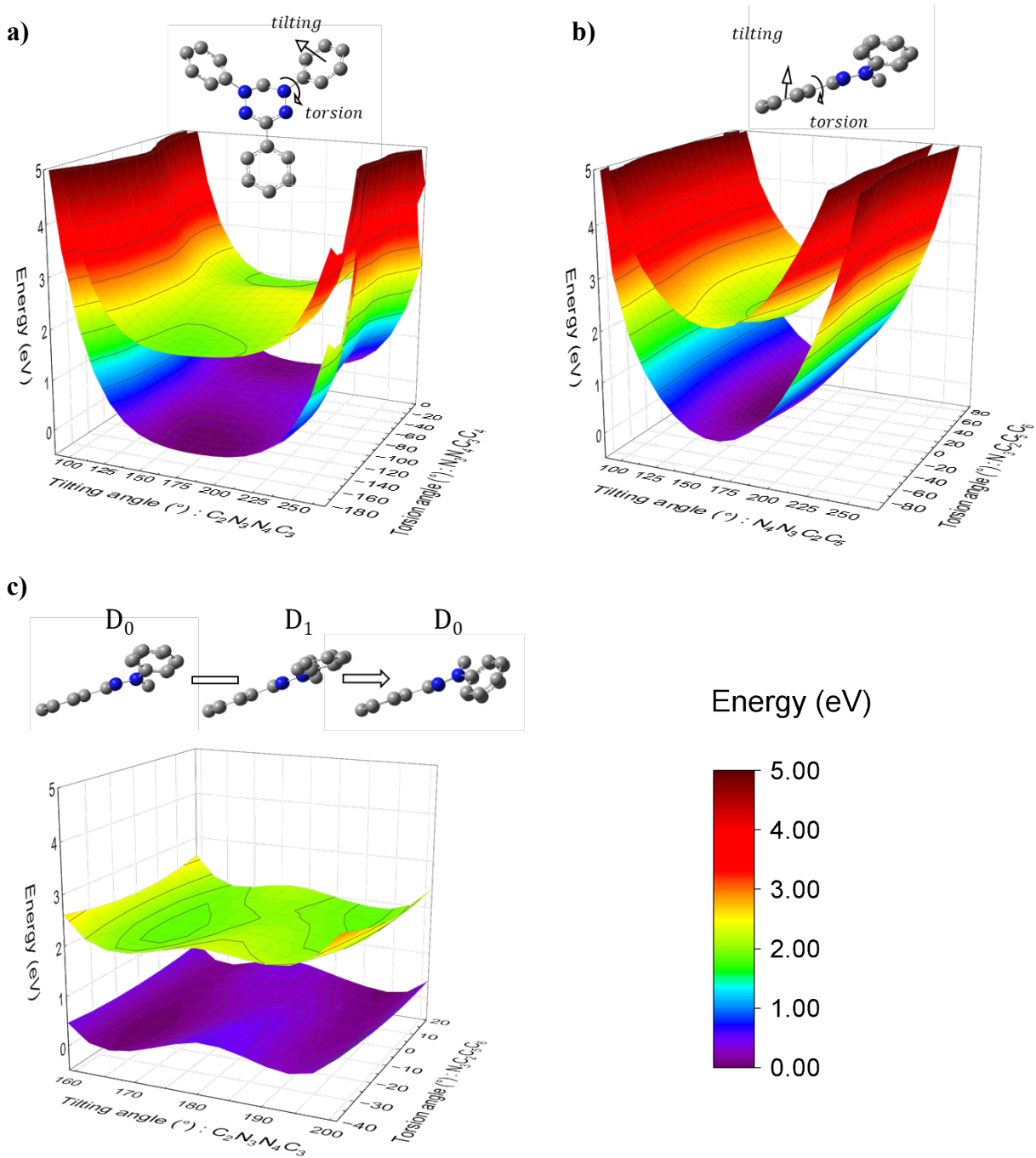

**Figure 1.** Potential energy surfaces (PES) of the $D_0$ and $D_1$ states along the torsion and tilting coordinates of the N-phenyl (a), C-phenyl (b) and both N-phenyl symmetrically (c) of TPV. The deformations applied are illustrated above the PES. Hydrogen atoms are omitted for clarity.

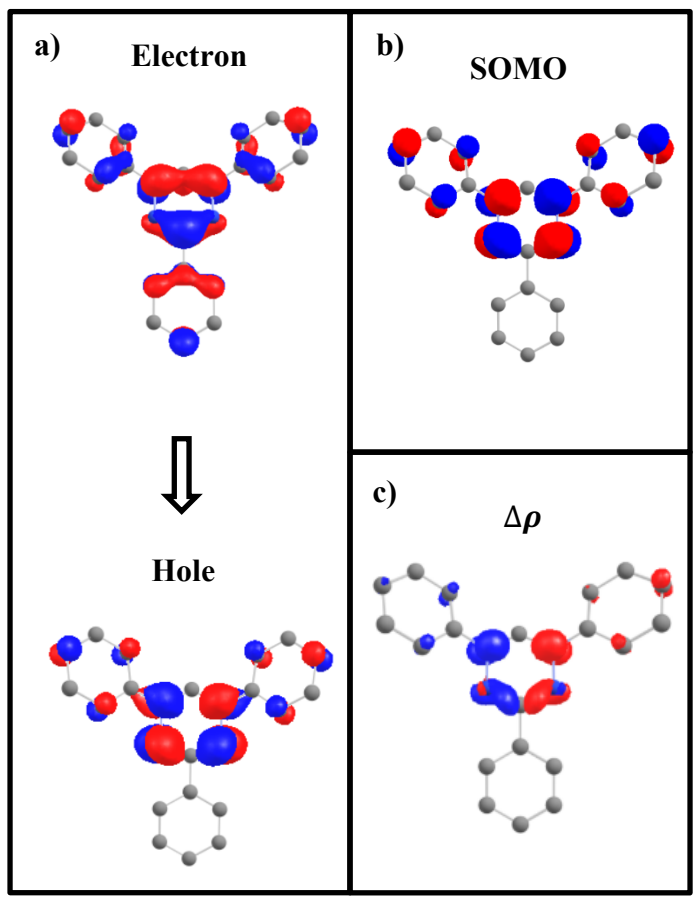

**Figure 2.** (a) Principal NTO (contribution is 94%) for the $D_0 \rightarrow D_1$ transition: electron (bottom) and hole (bottom) (is value: 0.04). (b) SOMO of the verdazyl radical (isovalue: 0.04). (c) Electron difference density map $\Delta\rho$ of the $D_0 \rightarrow D_1$ transition, showing the electronic redistribution upon excitation (isovalue: 0.004). Red and blue lobes represent positive and negative phases of the wavefunction (or density difference).

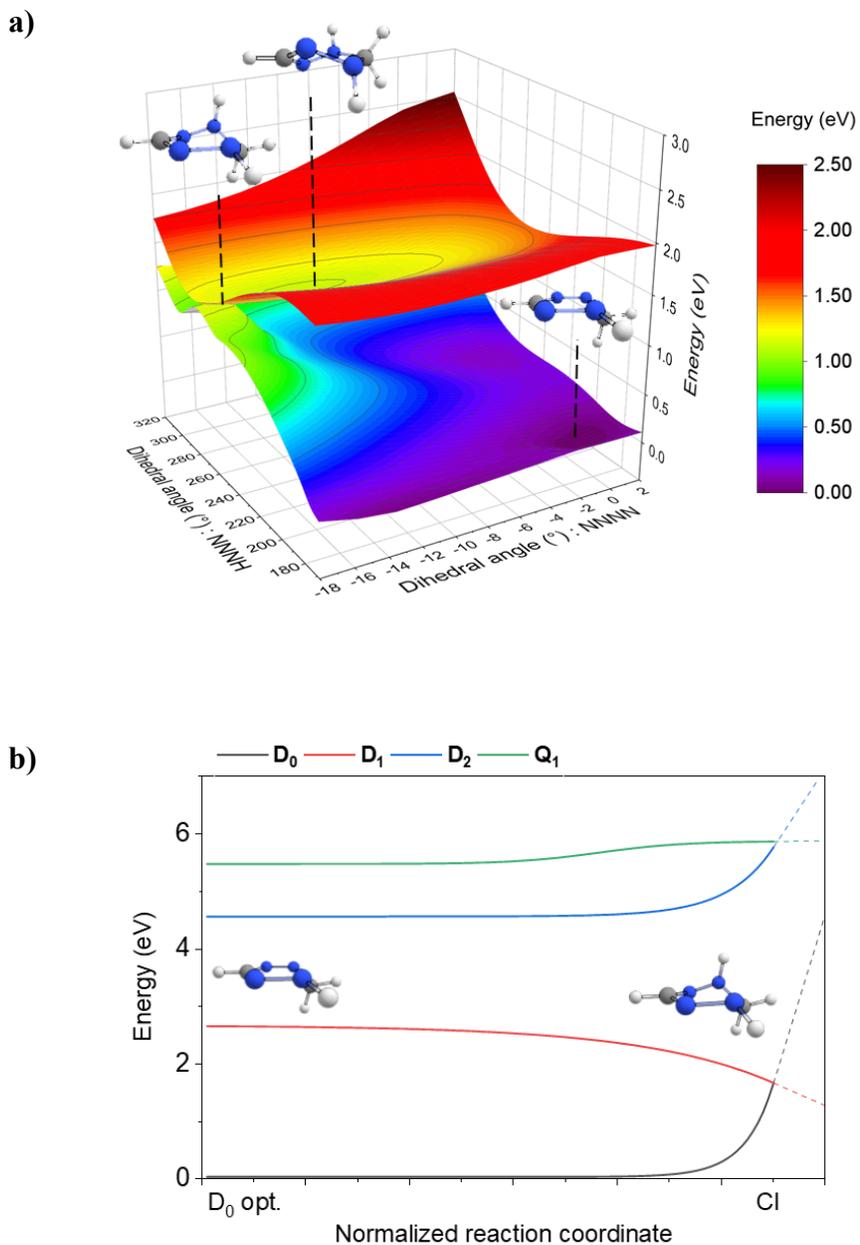

**Figure 3.** (a) Relaxed 3D PES scan of VR along $\widehat{NNNN}$ and $\widehat{NNNH}$, showing the $D_0$ and $D_1$ minima and the CI. (b) Minimum energy path (MEP) computed via nudged elastic band method between the $D_0$ minimum and the CI geometry of VR. The energies of the three excited states are also shown along the reaction coordinate.

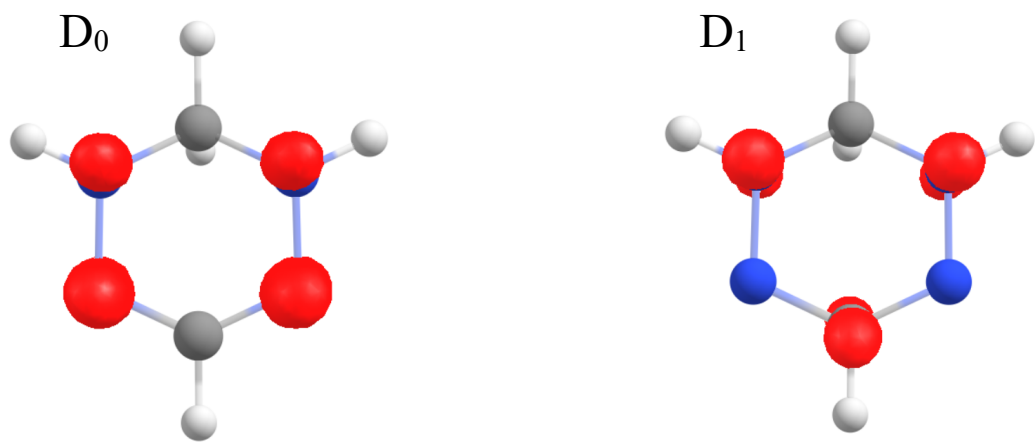

**Figure 4.** Electron spin density of $D_0$ (left) and $D_1$ (right) states of VR. Red lobes indicate positive spin density (isovalue: 0.005).

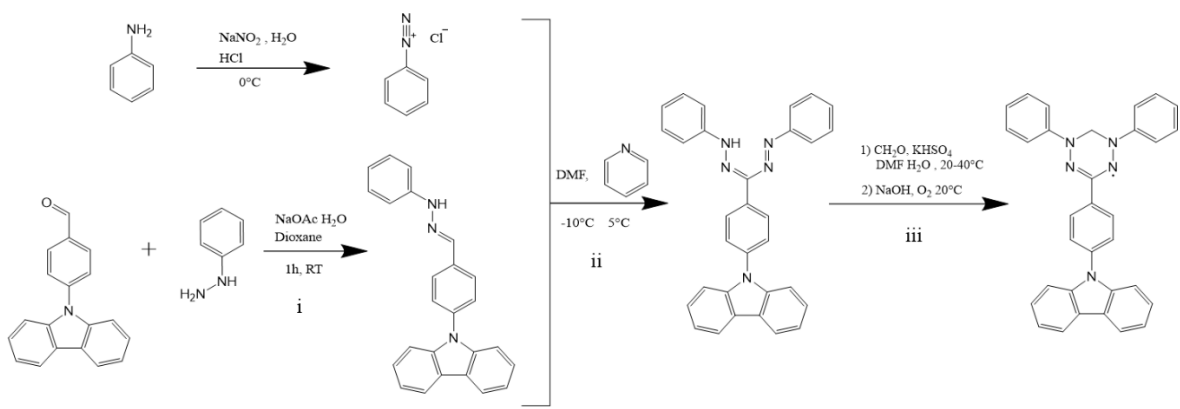

**Scheme 1.** Synthesis of TPV-Cz.

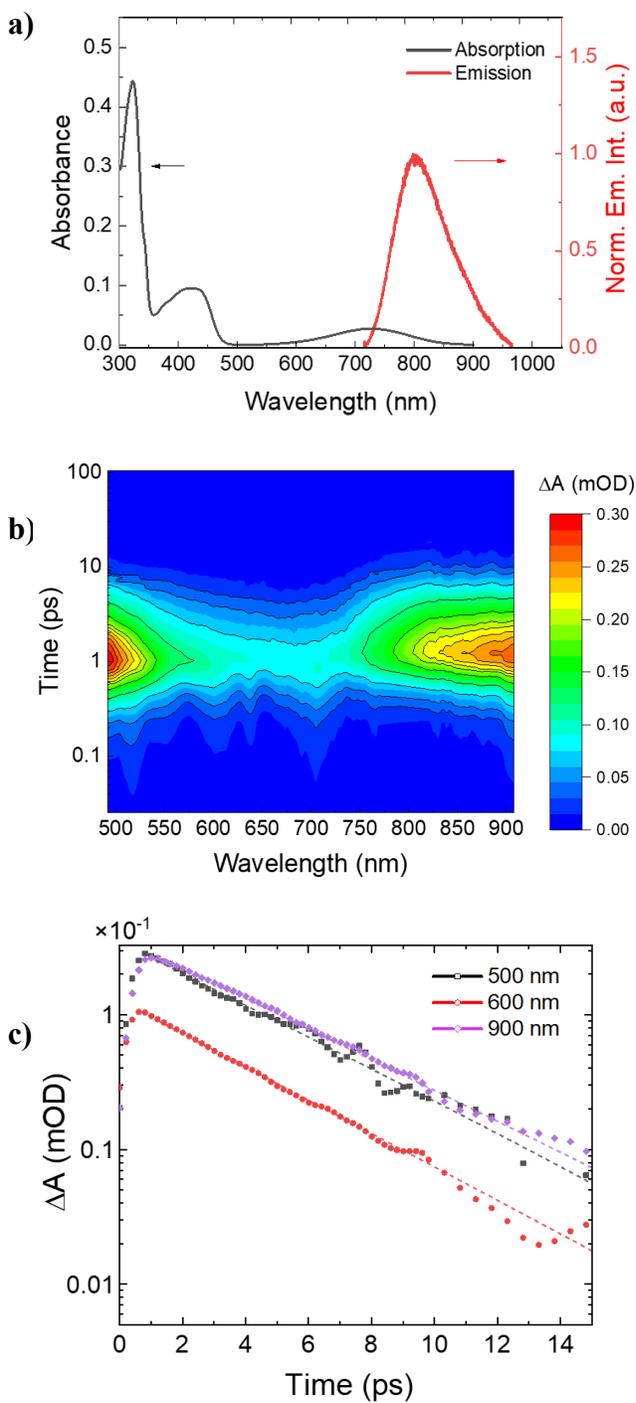

**Figure 5.** (a) Absorption spectrum in toluene and emission spectrum in PMMA (5 wt.%)(excitation at 710 nm) of TPV-Cz. (b) Transient absorption map and (c) kinetic traces with monoexponential fits at different wavelengths of TPV-Cz in toluene upon excitation at 750 nm.

|  | VR | TPV | TPV-Cz |
|---|---|---|---|
| $D_0 \rightarrow D_1$ (eV) | 2.65 | 1.88 | 1.88 |
| CI (eV) | 1.68 | 1.65 | 1.69 |
| $\widehat{NNNN}$ $D_0$ (°) | 0.0 | 0.0 | 0.0 |
| $\widehat{NNNN}$ CI (°) | -16.3 | -33.8 | -33.0 |
| $\widehat{NNNR}$ $D_0$ (°) | 168.1 | 161.5 | 160.6 |
| $\widehat{NNNR}$ CI (°) | 279.4 | 268.1 | 268.7 |

**Table 2.** Energies of the $D_0 \rightarrow D_1$ vertical transition and of the $D_0/D_1$ CI, along with the $\widehat{NNNN}$ and $\widehat{NNNR}$ dihedral angles in the optimized $D_0$ and CI geometries of the verdazyl ring, TPV, and TPV-Cz. R refers to H for the verdazyl ring and to C for TPV and TPV-Cz.

## ASSOCIATED CONTENT

**Supporting Information**.

Experimental procedures for the synthesis of 4-(4-(carbazol-9-yl)phenyl)-2,6-diphenylverdazyl (TPV-Cz); characterization data (NMR, HRMS, and EPR spectra); computational details including basis set and functional benchmarking; additional figures (potential energy surfaces, orbital analyses, photoluminescence and transient absorption data).

## AUTHOR INFORMATION

**Corresponding Author**

E-mail: s.kena-cohen@polymtl.ca

## ACKNOWLEDGMENTS

The authors gratefully acknowledge funding from the Natural Sciences and Engineering Research Council of Canada (NSERC) Discovery Grant and Alliance Quantum Consortium (Quantamole) programs. S.K.C. acknowledges support from the Canada Research Chairs Program. Computational work was supported by Calcul Québec and the Digital Research Alliance of Canada.

## ABBREVIATIONS

VR, verdazyl ring; TPV, 2,4,6-triphenylverdazyl; TPV-Cz ,4-(4-(carbazol-9-yl)phenyl)-2,6-diphenylverdazyl; DFT, density functional theory; TDDFT, time-dependent density functional theory; SF-TDDFT, spin-flip time-dependent density functional theory; CI, conical intersection; MEP, minimum energy path